\documentclass[a4paper,11pt]{article}
\usepackage{pos}
\usepackage[italicdiff]{physics}
\usepackage{cancel}
\usepackage{comment}
\usepackage{xcolor}
\usepackage{feynmp-auto}

\newcommand{\be}{\begin{equation}\begin{gathered}}
\newcommand{\ee}{\end{gathered}\end{equation}}

\newcommand{\bea}{\begin{equation}\begin{aligned}}
\newcommand{\eea}{\end{aligned}\end{equation}}

\newcommand{\bc}{\begin{cases}}
\newcommand{\ec}{\end{cases}}

\newcommand{\unit}[1]{%
    \ensuremath{\,\mathrm{#1}}%
}
\newcommand{\avg}[1]{%
    \ensuremath{\langle #1 \rangle}%
}

\newcounter{feynmancounter}
\newenvironment{feynman}[1]
  {
    \addtocounter{feynmancounter}{1}
    \begin{fmffile}{feynm\thefeynmancounter}
	\begin{fmfgraph*}(#1)
  }
  { 
    \end{fmfgraph*}
    \end{fmffile}
  }

\title{Effects of new physics in neutrino propagation}

\author[a,b]{J.M. Carmona}
\author[a,b]{J.L. Cort\'es}
\author[a,b,c]{J.J. Relancio}
\author*[a,b]{M.A. Reyes}

\affiliation[a]{Departamento de F\'{\i}sica Te\'orica,
Universidad de Zaragoza, Zaragoza 50009, Spain}

\affiliation[b]{Centro de Astropartículas y Física de Altas Energías (CAPA),
Universidad de Zaragoza, Zaragoza 50009}

\affiliation[c]{Departamento de Física, Universidad de Burgos, 09001 Burgos, Spain;}

\emailAdd{jcarmona@unizar.es}
\emailAdd{cortes@unizar.es}
\emailAdd{relancio@unizar.es}
\emailAdd{mkreyes@unizar.es}

\abstract{In this work we consider the effects of Lorentz Invariance Violation over the observed flux of very high-energy neutrinos. For that, we study the neutrino propagation in a Modified Dispersion Relation scenario with a superluminal velocity. This makes the neutrino unstable and causes a cut-off in the flux of detected neutrinos. Using simple models, one can approximate the location of the cut-off as a function of the parameters of new physics and the closest source.}

\FullConference{%
  Corfu Summer Institute 2021 "School and Workshops on Elementary Particle Physics and Gravity"\\
  29 August - 9 October 2021\\
  Corfu, Greece
}

\begin{document}
\maketitle

\section{Introduction}

General Relativity (GR) and Quantum Mechanics (QM) are the fundamental ingredients of our understanding of Nature. The former describes the gravitational interaction, and the latter is in the bases of the Standard Model (SM), which describes the other three interactions (electromagnetic, weak and strong). Nevertheless, every attempt of unification has encountered unsolvable difficulties. Specifically, one can notice the very different role that spacetime plays in each one. In GR, spacetime is a dynamic object that interacts with matter; however, in QM spacetime is a fixed frame described by Special Relativity (SR). So attempts of unification lead to change our conception of spacetime. Unfortunately, we do not have energetic enough experiments to give us hints about how to approach to a Quantum Gravity (QG) theory, but one natural way to proceed is to look for low energy signals of QG due to the dynamical nature of spacetime. One of these signals could be a breaking or deformation of the fundamental symmetry under SR, the Lorentz invariance. This leads to two different ways to go beyond the SM, the Lorentz Invariance Violation (LIV)~\cite{Colladay:1998fq,Kostelecky:2008ts} and the Doubly/Deformed Special Relativity (DSR) scenarios~\cite{AmelinoCamelia:2000mn}.

Every effect of LIV and DSR should be very small correction to the SM at the energies we have explored so far; however, they could be important and measurable at high enough energies~\cite{Mattingly:2005re,Liberati:2013xla,Addazi:2021xuf}. In addition, some models propose that these effects could be accumulative, so that they could become noticeable for particles traveling very large distances~\cite{Amelino-Camelia:1997ieq}. This encourages us to look for physics beyond the SM using observations of very high-energy extragalactic astroparticles, because they have energies orders of magnitude above the ones we can obtain in a laboratory, and also due to the fact that they involve propagation over very large distances, allowing us to observe possible accumulative effects.

Among the astrophysical messengers, neutrinos take a special role because they already contain signs of new physics beyond the SM, like their masses. But also because they are highly non-interacting, so they can travel very long distances without disappearing from the flux. So very high-energy neutrinos are ideal particles to study accumulative effects in LIV.

In this work we will consider the effects of LIV over the observed flux of very high-energy neutrinos. The main consequence is that the flux detected on Earth by IceCube~\cite{IceCube:2013cdw}, and by future experiments, would be modified with respect to the flux expected in SR. We start by introducing the LIV framework in Sec.~\ref{sec:LIV}. In Sec.~\ref{sec:ca} we describe this flux by a continuous evolution approximation. Nevertheless, this will not be enough to consider all the effects of LIV, so in Sec.~\ref{sec:ic} we will consider an instantaneous cascade approximation. Finally, in Sec.~\ref{sec:discussion} we will discuss some applications of these simple models.

\section{LIV framework}
\label{sec:LIV}

LIV effects at low energies must be small corrections with respect to the SM. Therefore, one can consider the LIV Lagrangian to be proportional to the inverse of some power $n$ of a new high-energy scale $\Lambda$, i.e. $\mathcal{L}_\text{LIV}\propto \Lambda^{-n}$, so that, 
for a great enough value of the scale $\Lambda$, we can always ensure that the corrections are small. In the literature it is usually considered $\Lambda \sim M_p$ (Planck mass).

Additionally, we can consider that LIV corrections affect only the free Lagrangian of neutrinos, so that the SM interaction Lagrangian remains the same, as well as the free Lagrangians of the other particles. Then, the only change with respect to the SM should be a Modified Dispersion Relation (MDR) for neutrinos, i.e.
\be 
    \abs{\vec p}= E \qty(1- \alpha \qty(\frac{E}{\Lambda})^n) \,,
\ee
with $\alpha=+1$ for a superluminal velocity or $\alpha=-1$ for a subluminal one. In this way we can use all the SM machinery for the computations of decay widths and cross sections by only taking into account the new MDR.

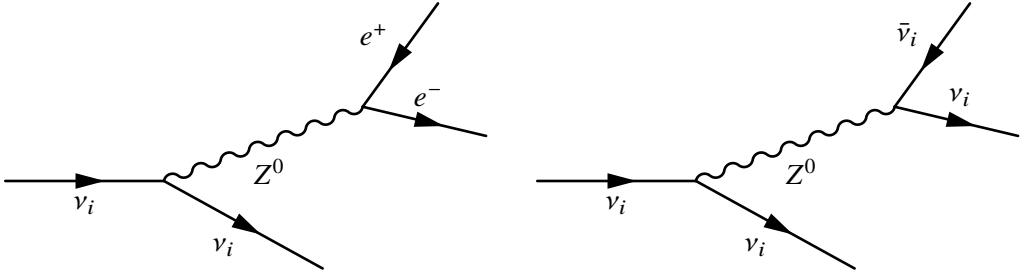
\begin{figure}[tb]
    \centering
    \begin{feynman}{180,100}
        \fmfleft{i1}
        \fmfright{o1,o2,o3}
        \fmf{fermion,label=$\nu_i$}{i1,v1}
        \fmf{fermion,label=$\nu_i$}{v1,o1}
        \fmf{boson,label=$Z^0$}{v1,v2}
        \fmf{fermion,label=$e^-$}{v2,o2}
        \fmf{fermion,label=$e^+$}{o3,v2}
        \fmfforce{(0.00w,0.33h)}{i1}
        \fmfforce{(0.33w,0.33h)}{v1}
        \fmfforce{(0.66w,0.00h)}{o1}
    \end{feynman}
    \begin{feynman}{180,100}
        \fmfleft{i1}
        \fmfright{o1,o2,o3}
        \fmf{fermion,label=$\nu_i$}{i1,v1}
        \fmf{fermion,label=$\nu_i$}{v1,o1}
        \fmf{boson,label=$Z^0$}{v1,v2}
        \fmf{fermion,label=$\nu_i$}{v2,o2}
        \fmf{fermion,label=$\bar\nu_i$}{o3,v2}
        \fmfforce{(0.00w,0.33h)}{i1}
        \fmfforce{(0.33w,0.33h)}{v1}
        \fmfforce{(0.66w,0.00h)}{o1}
    \end{feynman}
    \caption{Disintegration of neutrino through VPE (neutral channel) and NSpl}
    \label{fig:feynman_VPE_NSpl}
\end{figure}

A superluminal modification of the neutrino energy-momentum relation allows disintegrations that are forbidden in SM due to energy-momentum conservation. Specifically, now the neutrino is unstable and it can disintegrate through Vacuum (electron-positron) Pair Emission (VPE) and Neutrino Splitting (NSpl), processes depicted in Fig.~\ref{fig:feynman_VPE_NSpl}.

The VPE process consists of the disintegration of one neutrino into an electron-positron pair and a neutrino of lower energy. The pair can be produced through the emission of a boson $Z^0$ (neutral channel) for every flavour of the neutrino, or a boson $W^+$ (charged channel) only for electronic neutrinos. However, in this work we will only consider the neutral channel (relevant $5/6$ of the times)\footnote{For the simplest scenario of a flavor population of $(1:2:0)$ at the source, neutrino mixing makes each flavour appear approximately $1/3$ of the times. Only electron neutrinos can decay through the charged channel. If one assumes equal probabilities for the neutral and charged channels, then the charged channel is only relevant $1/6$ of the time.} to ease the computations. Due to the mass of the electron and positron of the final state, the process of VPE has an energy threshold, given by~\cite{Liberati:2013xla} 
\be 
    E_\text{th}^\text{VPE} \coloneq (4m_e^2 \Lambda^n)^{1/(n+2)} \,,
\ee
where $m_e$ is the electron mass. Additionally, using the Feynman rules one can compute the differential decay width\footnote{\label{foot:spin}This is the differential decay width without considering the average over the neutrino initial spin states. A factor $1/2$ is expected otherwise.} of VPE, obtaining~\cite{Carmona:2012tp}
\be
    \pdv{\Gamma^\text{VPE}(E,x)}{x} = \frac{32 E^5}{192\pi^3} G_F^2 \qty[(s_W^2-1/2)^2 + (s_W^2)^2] \qty(\frac{E}{\Lambda})^{3n} \;\frac{1}{3} (1-x^{n+1})^3 (1-x^{3}) \,,
    \label{eq:dif_width_VPE}
\ee
where $x$ is the fraction of the initial energy inherited by the final neutrino, $G_F$ is the Fermi coupling, and $s_W\coloneq \sin\theta_W$ is the sine of the Weinberg angle.

The NSpl process consists of the disintegration of a neutrino into two neutrinos and an antineutrino. Approximating neutrinos as massless, there is not any threshold for this process, and the differential decay width\footnote{See footnote \ref{foot:spin}} for the cases $n=1$ and $n=2$ can be written as\footnote{There exists a more general way to write the NSpl decay width for any $n$, but the energy fraction terms only factorize without mixing for $n=1$ and $n=2$.}~\cite{colineal_neu}
\bea
    \frac{\partial\Gamma^\text{NSpl}(E,x_1,x_2,x_3)}{\partial x_1 \partial x_2 \partial x_3} = \frac{32 E^5}{192\pi^3} G_F^2 \qty(\frac{E}{\Lambda})^{3n} &\frac{(n+1)^3}{4} \\ &\delta\qty(1-x_1-x_2-x_3) (1-x_1)^3 (1-x_2)^3 (1-x_3)^{3n-1} \,,
    \label{eq:dif_width_NSpl}
\eea
where $x_1$, $x_2$ and $x_3$ are the fractions of the initial energy inherited by the neutrino of the first vertex, the neutrino of the second vertex, and the antineutrino, respectively.

\section{Continuous evolution of the energy considering VPE}
\label{sec:ca}

We want to study how these two new disintegrations affect the flux of detected neutrinos. For that, let us first consider an individual neutrino traveling freely from an extragalactic source to our detector at Earth. In order to deal with cosmological distances, we are going to use the redshift $z$ as a way to follow the trajectory of the neutrino. Then, the neutrino is emitted at $z_e$ and detected at $z_d=0$. During this trip, the neutrino will suffer a continuous adiabatic loss of energy due to the expansion of the Universe. This energy loss can be characterized by a differential equation
\be
    \frac{1}{E}\frac{dE}{dz}=\frac{1}{1+z} \,.
    \label{eq:E_evol_classic}
\ee
This process is deterministic, so the energy of the neutrino is perfectly defined along the trajectory. If this is the only mechanism of energy loss, like in SR, its propagation is trivial and the relation between the emission and detection energies is given by
\be 
    E_e= (1+z_e) E_d\,.
    \label{eq:Ee_Ed_classic}
\ee

If we consider now the LIV framework presented in the last section, the neutrino can suffer additionally disintegrations through VPE and NSpl. For simplicity, let us consider only VPE in this section. Identifying the initial and final neutrino as the same, we can picture the VPE as a process of energy  loss due to the emission of an electron-positron  pairs. The disintegration is intrinsically stochastic; therefore, during the propagation, the neutrino has some probability of losing energy, which is proportional to the total decay width of VPE \cite{Carmona:2012tp},
\be
    \Gamma^\text{VPE}(E) = \frac{32 E^5}{192\pi^3} G_F^2 \qty[(s_W^2-1/2)^2 + (s_W^2)^2] \qty(\frac{E}{\Lambda})^{3n} \xi_n^\text{VPE}
    \,,
    \label{eq:total_width_VPE}
\ee
where $\xi_n^\text{VPE}$ is a constant dependent on $n$,
\bea
    \xi_n^\text{VPE} &\coloneq \frac{1}{3} \int_0^1 dx (1-x^{n+1})^3 (1-x^3) \\ &= \frac{1}{4} - \frac{3}{(n+2)(n+5)}+ \frac{3}{(2n+3)(2n+6)}-\frac{1}{(3n+4)(3n+7)} \,.
    \label{eq:xi_n_VPE}
\eea
Additionally, if there is a disintegration, the neutrino has a certain probability distribution for its final fraction of energy, given by
\be
    \mathcal{P}^\text{VPE}(x) \coloneq \frac{1}{ \Gamma^\text{VPE}(E)}  \pdv{\Gamma^\text{VPE}(E,x)}{x} = \frac{1}{\xi_n^\text{VPE}} \; \frac{1}{3} (1-x^{n+1})^3 (1-x^3) \,.
    \label{eq:prob_VPE}
\ee
Then, the energy of each individual neutrino becomes undetermined during its propagation making impossible to follow its evolution. However, if instead of considering the propagation of one neutrino we consider a large enough number of them, these probabilities should allow us to study the typical (mean) evolution of the energy along the trajectory. The differential variation of the energy $dE$ in a differential time $dt$ is given by the product of the probability of disintegration, $dt\,\Gamma^\text{VPE}(E)$, and the mean energy loss, $-E\avg{1-x}$. Then
\be
    \frac{dE}{dt} = - E\, \Gamma^\text{VPE}(E) \avg{1-x} \,.
    \label{eq:vpe_rate}
\ee
One can solve the previous equation and write the variation in terms of the redshift using the relation between $dt$ and $dz$ given by
\be
    \frac{dt}{dz}=\frac{-1}{(1+z)H(z)}=\frac{-1}{(1+z)H_0\sqrt{(1+z)^3 \Omega_m+ \Omega_\Lambda}} \,,
    \label{eq:dt_dz}
\ee
with $H(z)$ the Hubble parameter, defined by the Hubble constant $H_0$ and the density fractions of matter and dark energy, $\Omega_m$ and $\Omega_\Lambda$, respectively. Hence, the evolution of the energy due to VPE is given by
\be
    \frac{1}{E}\frac{dE}{dz} = \frac{\Gamma^\text{VPE}(E)}{(1+z)H(z)} \avg{1-x} \equiv \qty(\frac{E}{E_r^\text{VPE}})^{5+3n} \frac{\avg{1-x}}{(1+z)\sqrt{(1+z)^3 \Omega_m+ \Omega_\Lambda}} \,,
    \label{eq:E_evol_VPE}
\ee
where we have defined
\be 
    E_r^\text{VPE} \coloneq \qty(\frac{32}{192\pi^3} \frac{G_F^2}{H_0 \Lambda^{3n}} \qty[(s_W^2-1/2)^2 + (s_W^2)^2] \xi_n^\text{VPE})^{-1/(5+3n)} \,.
    \label{eq:E_r_VPE}
\ee
From the probability distribution of Eq.~\eqref{eq:prob_VPE} we can compute $\avg{1-x}$, which can be written as
\be
    \avg{1-x} = \frac{1}{\xi_n^\text{VPE}} \; \frac{1}{3} \int_0^1 dx (1-x^{n+1})^3 (1-x^3) (1-x) \equiv \frac{{\xi'_n}^\text{VPE}}{\xi_n^\text{VPE}}
    \,,
\ee
with
\bea
    {\xi'_n}^\text{VPE} &\coloneq \frac{1}{3} \int_0^1 dx (1-x^{n+1})^3 (1-x^3) (1-x) \\ &= \frac{3}{20}-\frac{6(n+4)}{(n+2)(n+3)(n+5)(n+6)} +\frac{6(2n+5)}{(2n+3)(2n+4)(2n+6)(2n+7)} \\& \hspace{15em} -\frac{2(3n+6)}{(3n+4)(3n+5)(3n+7)(3n+8)}
    \,.
\eea

Let us notice that the energy $E_r^\text{VPE}$ acts as an effective threshold for the VPE, because if the energy is above the energy scale, then
\be 
    \qty(\frac{E}{E_r^\text{VPE}})^{5+3n} = \frac{\Gamma^\text{VPE}(E)}{H_0} \gg 1 \,,
    \label{eq:vpe_strong}
\ee
because of the large power $5+3n$ of the quotient of energies, and so the energy loss in Eq.~\eqref{eq:E_evol_VPE} will be very strong. We can call this the strong regime. On the contrary, if the energy is below the effective threshold, the energy loss by VPE is negligible, recovering Eq.~\eqref{eq:E_evol_classic}. We can call this the standard regime. In this way, the process of VPE has two different thresholds: a real one, $E_\text{th}^\text{VPE}$, and an effective one, $E_r^\text{VPE}$. The greater of the two will control where the VPE stops, but the relative magnitude between both is a function of the scale of new physics and the order of the correction,
\be 
    \qty(E_\text{th}^\text{VPE}/E_r^\text{VPE})^{5+3n} \equiv \qty(\Lambda^0_n/\Lambda)^{n/(n+2)} \,,
\ee
with $\Lambda^0_n$ a constant given by
\be 
    \Lambda^0_n \coloneq (4m_e^2)^{(5+3n)/n} \qty(\frac{32}{192\pi^3} \frac{G_F}{H_0}\qty[(s_W^2-1/2)^2 + (s_W^2)^2] \xi_n^\text{VPE})^{(2+n)/n} \,.
\ee
We are interested in linear and quadratic corrections, i.e. $n=1$ and $n=2$ respectively, and values of $\Lambda$ below or equal to the Planck scale.
\begin{figure}[tb]
    \centering
    \includegraphics[width=0.6\textwidth]{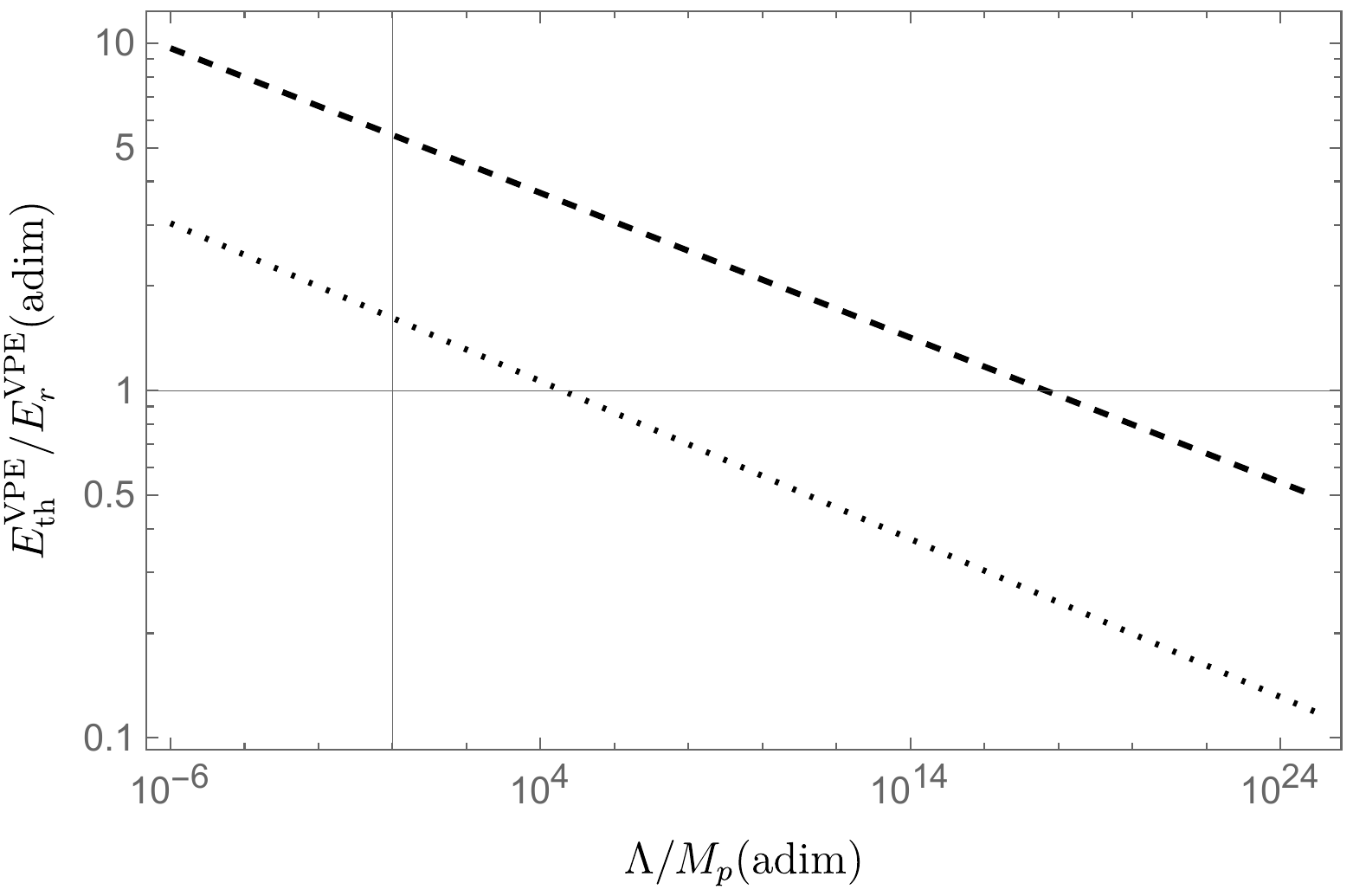}
    \caption{Quotient between energy thresholds for $n=1$ (dashed) and $n=2$ (dotted)}
    \label{fig:comparison_Eth_Er}
\end{figure}
As we see in Fig.~\ref{fig:comparison_Eth_Er}, 
under these conditions, $E_\text{th}^\text{VPE}$ is always greater than $E_r^\text{VPE}$. This implies that the real threshold is the relevant scale to study where the VPE stops, but additionally, it implies that the energy loss is always in the strong regime because, for every neutrino suffering VPE, its energy will fulfill  $E\ge E_\text{th}^\text{VPE}> E_r^\text{VPE}$.

Let us consider neutrinos emitted with energy above the threshold, so they can produce VPE. Combining the VPE energy loss, Eq.~\eqref{eq:E_evol_VPE}, with the standard adiabatic energy loss, Eq.~\eqref{eq:E_evol_classic}, we can get a differential equation for the evolution of the energy of a neutrino emitted with energy above the VPE threshold~\cite{Carmona:2019xxp},
\be 
    \frac{1}{E} \frac{dE}{dz} = \frac{1}{1+z} + \qty(\frac{E}{E_r^\text{VPE}})^{5+3n} \frac{\avg{1-x}}{(1+z)\sqrt{(1+z)^3 \Omega_m + \Omega_\Lambda}} \,.
    \label{eq:E_evol}
\ee
This equation will rule the evolution of the energy until the neutrinos reach the threshold energy or are detected. In the last case, one can integrate\footnote{To integrate Eq.~\eqref{eq:E_evol} one can do a change of variables from $E$ to the redshift-invariant variable $\tilde E=E/(1+z)$.} Eq.~\eqref{eq:E_evol} from the emission until the detection, hence obtaining that the relation between the emission and detection energy is
\be 
    E_e = (1+z_e) E_d \qty[1-\qty(\frac{E_d}{E_r^\text{VPE}})^{5+3n} \avg{1-x}\,J_n\qty(z_e,0)]^{-1/(5+3n)}\,, \quad (E_d\ge E_\text{th}^\text{VPE})\,,
    \label{eq:Ee_Ed_th}
\ee
where $J_n(z_i,z_f)$ is a shorthand for
\be 
    J_n(z_i,z_f) \coloneqq \int_{z_f}^{z_i} dz \frac{(1+z)^{4+3n}}{\sqrt{(1+z)^3 \Omega_m + \Omega_\Lambda}} \,.
\ee

Let us notice that, contrary to the trivial case shown in Eq.~\eqref{eq:Ee_Ed_classic}, now the equation relating emission and detection does not always have solution for every combination of the energy of emission, detection, and location of the source. In fact, we are restricted to the parameters that make the term inside of the square brackets larger than zero, i.e. 
\be
     J_n(z_e,0) < \frac{1}{\avg{1-x}} \, \qty(\frac{E_r^\text{VPE}}{E_d})^{5+3n} \ll 1\,, \quad (E_d\ge E_\text{th}^\text{VPE}>E_r^\text{VPE}) \,.
\ee
Then we expect $z_e\approx 0$ for every energy of detection greater or equal to the VPE threshold, and we see that the number of sources capable to contribute to the detected flux of energy $E_d$ tends to zero. We do not expect to detect neutrinos above $E_\text{th}^\text{VPE}$.

If we now consider neutrinos detected with energy below the threshold, Eq.~\eqref{eq:E_evol} will rule the evolution of their energy until the neutrinos reach the threshold energy at some point $z_\text{th}^\text{VPE}$ of the trajectory. From that point the evolution is trivial, so we can solve $z_\text{th}^\text{VPE}$ as a function of the detection energy, 
\be 
    z_\text{th}^\text{VPE} (E_d) \coloneq E_\text{th}^\text{VPE}/E_d-1 \,, \quad (E_d\le E_\text{th}^\text{VPE})\,.
\ee
Integrating Eq.~\eqref{eq:E_evol} from the emission to $z_\text{th}^\text{VPE}(E_d)$, we get that the relation between the emission and detection energy is now
\be 
    E_e = (1+z_e) E_d \qty[1-\qty(\frac{E_d}{E_r^\text{VPE}})^{5+3n} \avg{1-x}\,J_n\qty(z_e,E_\text{th}^\text{VPE}/E_d-1)]^{-1/(5+3n)} \,, \quad (E_d\le E_\text{th}^\text{VPE})\,.
    \label{eq:Ee_Ed_th_2}
\ee
One more time, the existence of solution of the previous equation imposes
\be
     J_n\qty(z_e,E_\text{th}^\text{VPE}/E_d-1) < \frac{1}{\avg{1-x}} \, \qty(\frac{E_r^\text{VPE}}{E_d})^{5+3n}\,, \quad (E_d\le E_\text{th}^\text{VPE})\,,
     \label{eq:condition}
\ee
but in contrast to the previous case, we now have to possibilities: $(E_\text{th}^\text{VPE}>E_d>E_r^\text{VPE})$ or  $(E_\text{th}^\text{VPE}>E_r^\text{VPE}>E_d)$. In the last case, the condition \eqref{eq:condition} is easily satisfied without imposing any condition over $z_e$, so one can expect to detect neutrinos below $E_r^\text{VPE}$. However, the most interesting case is to consider the most energetic neutrinos below the threshold, i.e. $(E_\text{th}^\text{VPE}>E_d>E_r^\text{VPE})$, because in that case Eq.~\eqref{eq:Ee_Ed_th_2} only has solution if $z_e$ is below some $z_c(E_d)$, given by
\begin{equation}\begin{split}
    z_c(E_d) \coloneq  z_\text{th}^\text{VPE} (E_d) +  \qty(\frac{E_r^\text{VPE}}{E_\text{th}^\text{VPE}})^{4+3n} \qty(\frac{E_r^\text{VPE}}{E_d}) \, \frac{1}{\avg{1-x}}\, \sqrt{(1+z_\text{th}^\text{VPE}(E_d))^3 \Omega_m + \Omega_\Lambda} \,,
    \\ (E_\text{th}^\text{VPE}>E_d>E_r^\text{VPE}) \,.
\end{split}\end{equation}
Let us notice that $z_c(E_d)$ is a monotonic decreasing function on $E_d$. Then, there exists some energy $E_\text{cut}$ for which $z_c(E_\text{cut})$ will coincide with the closest source $z_\text{min}$. This means that we are not going to find any neutrino with energy above $E_\text{cut}$, because this would require a source closer than the closest source. Recalling that for linear and quadratic corrections, and $\Lambda$ below or equal the Planck scale, $E_\text{th}^\text{VPE}$ is always greater than $E_r^\text{VPE}$, we can expect $z_c(E_d)\approx z_\text{th}^\text{VPE}(E_d)$, so we can conclude that considering VPE implies a cut-off in the flux of detected neutrinos and it is located at an energy $E_\text{cut}$ given approximately by
\be 
    E_\text{cut} \approx E_\text{th}^\text{VPE}/(1+z_\text{min}) \,.
\ee
Using that the threshold energy is a function of $\Lambda$ and $n$, we can write the cut-off energy as
\be 
    E_\text{cut} \approx \frac{\Lambda^{n/(n+2)}}{1+z_\text{min}} \; (2m_e)^{2/(n+2)} \,.
\ee
This allows us to predict approximately the location of the cut-off as a function of the parameters of the new physics. But this relation can also be written in the other way around, so that we can use it to put constraints in the possible values of $\Lambda$ given an observation of neutrinos of energy $E_d$, 
\be 
    \Lambda > \qty(E_d (1+z_\text{min}))^{(n+2)/n} (2m_e)^{-2/n} \,.
\ee

\section{Instantaneous cascade approximation for the flux considering VPE and NSpl}
\label{sec:ic}

In the previous section we have seen that, due to the peculiar energy dependence of the VPE decay width, particles emitted with energies above $E_\text{th}^\text{VPE}$ reach the threshold energy almost instantaneously and are detected with energy $E_d\approx E_\text{th}^\text{VPE}/(1+z_e)$. As the NSpl decay width has the same energy dependence, this encourages us to approximate both VPE and NSpl as instantaneous effects. For that, we are again interested in a regime in which the effects are strong, i.e. $\Gamma^\text{VPE}(E)\gg H_0$ and $\Gamma^\text{NSpl}(E)\gg H_0$. The condition for the strong regime for VPE has already been studied in the previous section, Eq.~\eqref{eq:vpe_strong}, and for NSpl it can be done in the same way. Thence,
\be 
    \frac{\Gamma^\text{NSpl}(E)}{H_0} = \qty(\frac{E}{E_r^\text{NSpl}})^{5+3n} \gg 1 \,,
    \label{eq:nspl_strong}
\ee
with $\xi_n^\text{NSpl}$ given by~\cite{colineal_neu}
\bea
    \xi_n^\text{NSpl}  &\coloneq \frac{(n+1)^3}{4} \int_0^1 dx_1 \int_0^{1-x_1}  dx_2 \int_0^{1-x_1-x_2} dx_3 \, \delta\qty(1-x_1-x_2-x_3)\, (1-x_1)^3 (1-x_2)^3 (1-x_3)^{3n-1} \\
     &= \frac{(n+1)^3}{4} \,\bigg[\frac{1}{(3n+1)} - \frac{3}{(3n+2)} + \frac{7}{2(3n+3)} - \frac{2}{(3n+4)} + \frac{3}{5(3n+5)} - \frac{1}{10(3n+6)} \\ &\hspace{28em} + \frac{1}{140(3n+7)}\bigg]\,,
\eea
and where we have also defined an effective threshold for NSpl, similarly as the case of VPE,
\be 
    E_r^\text{NSpl} \coloneq \qty(\frac{32}{192\pi^3} \frac{G_F^2}{H_0 \Lambda^{3n}}\, \xi_n^\text{NSpl})^{-1/(5+3n)} \,.
    \label{eq:E_r_NSpl}
\ee
So we obtain that for NSpl, similarly to VPE, if the energy of emission is above the effective threshold the effect can be considered as instantaneous. In this case, in contrast to VPE, it does not exist a real threshold, but $E_r^\text{NSpl}$ will control when the NSpl stops. It is also worth to notice that dividing Eqs.~\eqref{eq:E_r_VPE} and \eqref{eq:E_r_NSpl}, one can check that $E_r^\text{VPE}/E_r^\text{NSpl} \sim 1$, so both effective thresholds are very similar and then also $E_\text{th}^\text{VPE}$ is greater than $E_r^\text{NSpl}$.

Let us assume that $N_e$ neutrinos were emitted with very high-energy such that $E_e>E_\text{th}^\text{VPE}>E_r^\text{NSpl}$; then, they will be able to produce both effects, NSpl and VPE. Specifically, they will disintegrate through  both kinds of processes until their energies reach the VPE threshold, and then, they will produce only NSpl until reaching the NSpl effective threshold. Let us assume that in the first part of the trajectory each neutrino suffers $p$ VPE and $q$ NSpl disintegrations. Then, we expect that 
\be 
    E_\text{th}^\text{VPE} = E_e\; \Big(\avg{x}^\text{VPE}\Big)^p \Big(\avg{x}^\text{NSpl}\Big)^q \,,
    \label{eq:before_q}
\ee
where $\avg{x}^\text{VPE}$ and $\avg{x}^\text{NSpl}$ are the average fraction of energy of a secondary neutrino after a disintegration through VPE and NSpl, respectively. But we also know that the the number of disintegrations through each process is proportional to their decay width. Thence, for enough disintegrations we also expect
\be 
    p/q \approx \frac{\Gamma^\text{VPE}(E)}{\Gamma^\text{NSpl}(E)} = \qty(\frac{E_r^\text{NSpl}}{E_r^\text{VPE}})^{5+3n} \eqcolon r \,.
\ee
Using that $p=qr$, we can solve Eq.~\eqref{eq:before_q} for $q$, obtaining
\be 
    q = \frac{\ln(E_e/E_\text{th}^\text{VPE})}{-\ln(\Big(\avg{x}^\text{VPE}\Big)^r \avg{x}^\text{NSpl})} \,.
\ee
Now, once the neutrino has reached the VPE threshold, it will continue losing energy through NSpl until reaching $E_r^\text{NSpl}$. Assuming there has been $q'$ disintegrations, we get
\be 
    E_r^\text{NSpl} = E_\text{th}^\text{VPE} \; \Big(\avg{x}^\text{NSpl}\Big)^{q'} \quad\rightarrow\quad q' = \frac{\ln(E_\text{th}^\text{VPE}/E_r^\text{NSpl})}{-\ln(\avg{x}^\text{NSpl})} \,.
\ee
Under the instantaneous approximation, the NSpl effective threshold will be reached very close to the source. In this way, we can consider an instantaneous cascade in which the $N_e$ neutrinos emitted with energies $E_e>E_\text{th}^\text{VPE}>E_r^\text{NSpl}$ will instantaneously produce $3^q\,3^{q'}\, N_e$ neutrinos of energy $E_r^\text{NSpl}$. If instead the $N_e$ neutrinos were emitted with energies $E_\text{th}^\text{VPE}>E_e>E_r^\text{NSpl}$, using the same procedure as before, we get that the $N_e$ neutrinos can be substituted with $3^{q''}\, N_e$ neutrinos of energy $E_r^\text{NSpl}$, being
\be 
    q'' = \frac{\ln(E_e/E_r^\text{NSpl})}{-\ln(\avg{x}^\text{NSpl})} \,.
\ee

In any case, all the neutrinos emitted above the NSpl effective threshold will produce a bunch of neutrinos that will propagate as if they were emitted from the source at $z_e$ with energy $E_r^\text{NSpl}$, so we do not expect to detect neutrinos above 
\be 
    E_\text{cut} \approx \frac{E_r^\text{NSpl}}{1+z_\text{min}} =  \frac{\Lambda^{3n/(5+3n)}}{1+z_\text{min}} \qty(\frac{32}{192\pi^3} \frac{G_F^2}{H_0}\,\xi_n^\text{NSpl})^{-1/(5+3n)} \,,
\ee
or written the other way around, we get a bound on $\Lambda$,
\be 
    \Lambda > \qty(E_d (1+z_\text{min}))^{(5+3n)/(3n)} \qty(\frac{32}{192\pi^3} \frac{G_F^2}{H_0}\,\xi_n^\text{NSpl})^{1/3n} \,,
\ee
from the observation of neutrinos with an energy $E_d$.

\section{Discussion}
\label{sec:discussion}

We have seen that a superluminal velocity due to a MDR makes neutrino unstable above certain threshold energies, allowing it to disintegrate through VPE and NSpl. In consequence, we expect a cut-off in the detected spectrum of neutrinos, and, using simple models, we have made an estimation of the energy of the cut-off as a function of the values of the new physics parameters $\Lambda$ and $n$, and the distance to the closest source, $z_\text{min}$. It is interesting to note that a cut-off in the neutrino spectrum originated from LIV would not only directly point to the LIV parameters ($\Lambda$ and $n$), but also to $z_\text{min}$, being informative on the distribution of the sources. This is a consequence of the instantaneous approximation, which, as we have seen in this study, is indeed a generic characeristic of the LIV processes affecting the propagation of very high-energy neutrinos.

The estimation of the location of the cut-off can be used to constrain the values of $\Lambda$ compatible with current observations of very high-energy neutrinos. In particular, one could pay special attention to the observations of IceCube, where one can see a lack of events above a few \unit{PeV}~\cite{IceCube:2013cdw}. This is specially striking if one takes into account that at $6.3\unit{PeV}$ one would expect to observe the Glashow resonance~\cite{Glashow:1960zz} (production of a non-virtual $W^-$ boson by the interaction of electron antineutrinos), boosting the probability of detection at that energy.

Before 2016, the most energetic detected neutrino was around $2\unit{PeV}$, and some works~\cite{Stecker:2014oxa} proposed that this could be a consequence of a cut-off in the detected spectrum caused by LIV disintegrations, so they performed Monte Carlo simulations considering VPE and NSpl for different values of $\Lambda$, obtaining that for a detection event around $2\unit{PeV}$ (with $n=2$, considering sources from $z=0.5$ to $z=2$), one can expect $\Lambda>1.4\cdot 10^{11}\unit{PeV}\approx 0.011 M_P$. Instead of performing a simulation, we can now apply the simple model discussed here with the same conditions ($n=2$, $E_d=2\unit{PeV}$, $z_\text{min}=0.5$), and we obtain $\Lambda>3.6\cdot 10^{10}\unit{PeV}\approx 0.003 M_P$. While the simulation gives a more restrictive constraint, the simple model can provide a estimation of the order of magnitude.

On 2016, an event compatible with the Glashow resonance was reported by IceCube~\cite{IceCube:2021rpz}, showing that there is not a cut-off at $2\unit{PeV}$. In order to improve the constraint over $\Lambda$, new Monte Carlo simulations have to be done; however, using the simple models, one can get a direct estimation, obtaining that for $n=2$, $E_d=6.3\unit{PeV}$ and $z_\text{min}=0.5$, the value of $\Lambda$ must be above $3.0\cdot 10^{11}\unit{PeV}\approx 0.024 M_P$.

One can argue that if $\Lambda$ is very large, $\Lambda\sim M_P$, we are still not sensitive to that new physics, and the apparent cut-off in the spectrum is due the fact that the sensitivity of the detector is restricted to energies below $10\unit{PeV}$. In that case, for $\Lambda= M_P$, one can use the simple models to estimate a location of the cut-off, which for $n=2$ in the less restrictive case ($z_\text{min}=0$) is around $E_\text{cut}\approx 72\unit{PeV}$. The next generation of detectors, like IceCube-Gen2~\cite{IceCube:2019pna}, aim to detect neutrinos with energies up to $\unit{EeV}=10^3\unit{PeV}$, so we expect that in the very near future the detectors will be sensitive to signatures that stem from $\Lambda$ around the Planck scale. 

Additionally, it could be interesting to identify a LIV signature, not only from the cut-off, but through the whole energy dependence (shape) of the detected neutrino flux. Nevertheless, a prediction of the flux energy dependence is beyond these simple models, requiring a Monte Carlo simulation to identify any modification due to LIV.

\bibliographystyle{JHEP}
\bibliography{QuGraPhenoBib,unpublished}



\end{document}